\documentclass[prb, preprint, showpacs]{revtex4}
\usepackage{graphicx,epsfig}
\usepackage{bm}
\usepackage{dcolumn}

\def\bea{\begin{eqnarray}}
\def\eea{\end{eqnarray}}
\def\be{\begin{equation}}
\def\ee{\end{equation}}

\def\degree{$^{\circ}$}

\begin{document}
\title{ Experimentally Constrained Molecular Relaxation: The case of hydrogenated amorphous silicon}

\author{Parthapratim Biswas}
\email{Partha.Biswas@usm.edu}
\affiliation{Department of Physics and Astronomy, The University of Southern Mississippi, 
Hattiesburg, MS 39401}

\author{Raymond Atta-Fynn}
\email{attafynn@uta.edu}
\affiliation{Department of Physics and Astronomy, The University of Texas at 
Arlington, Arlington, TX 76019}

\author{D.~A.~Drabold}
\email{drabold@ohio.edu}
\affiliation{Department of Physics and Astronomy, Ohio University, 
Athens, OH 45701 }


\begin{abstract}
We have extended our experimentally constrained molecular relaxation
technique (P.~Biswas {\it et al}, Phys.~Rev.~B {\bf 71} 54204 (2005)) 
to hydrogenated amorphous silicon: a 540-atom model with 7.4\% 
hydrogen and a 611-atom model  with  22\%  hydrogen were constructed.
Starting from a random configuration, using physically relevant constraints, 
{\it ab initio} interactions and the experimental static structure factor, 
we  construct realistic models of hydrogenated amorphous 
silicon. 
Our models confirm
the presence of a high frequency localized band in the vibrational density 
of states due to Si-H vibration that has been observed in a recent vibrational 
transient grating measurements on plasma enhanced chemical vapor deposited 
films of hydrogenated amorphous silicon. 

\end{abstract}

\maketitle

Ideal models of materials must be consistent with known experimentally-determined attributes 
{\it and} must also be near a minimum of an accurate energy functional. ``Information-based" methods 
such as the reverse Monte Carlo (RMC) method\cite{gere,McG,Biswas} produce models in agreement 
with experiment (typically a structural measurement), though often discrepant with chemical and 
bonding considerations, while molecular dynamics (MD) simulations suffer from artifacts of 
over-rapid quenches and small system sizes. To jointly satisfy these approaches, we introduced 
the Experimentally Constrained Molecular Relaxation (ECMR) method for amorphous materials and 
applied it with success to the complex binary glass GeSe$_2$~\cite{ecmr1}. Intriguing related ideas based upon Bayesian methods are emerging in 
studies of biomolecules\cite{bayes}, and our scheme is also reminiscent of the empirical potential structure 
refinement (EPSR) approach\cite{soper}.

In this paper, we extend this approach to hydrogenated amorphous silicon  (a-Si:H) a key electronic material
with myriad applications and a great many fundamental puzzles associated with structure, hydrogen dynamics
and properties under illumination\cite{Street}.  Owing  to  the complex nature of bonding between Si and H, a
straightforward application of Monte Carlo or MD simulation 
encounters difficulty in producing experimentally realistic models of hydrogenated amorphous silicon. 

Conventional  approaches to modeling a-Si:H can be divided 
into two categories: {\em static} and {\em dynamic}. In the static 
approach~\cite{Normand, Holender, Tuttle, Min}  a set of  predetermined  
geometric and chemical rules are employed to minimize the total 
energy either by Monte Carlo or some other minimization scheme, such as, 
conjugate gradient method. The so-called Wooten, Winer and Weaire 
(WWW)~\cite{Wooten} algorithm is the classic example of such a technique, 
and uses specific  Monte Carlo moves (the WWW  {\it bond-switch}) 
to produce excellent continuous random network (CRN) model of amorphous 
silicon. The network is then hydrogenated after creating
dangling-bond defects (threefold sites) and passivating the dangling bond with H,  or by breaking 
a bond between a  five-fold and four-fold coordinated atom and again repairing the threefold site with H with a suitable structural relaxation. While this 
approach produces models that exhibit good agreement with 
experiments, some of the troubling issues that warrant further 
investigation include distribution of hydrogen atoms and the amount of 
strain present in the  network. In the dynamic approach~\cite{Drabold, Buda, Klein} 
a mixture of silicon  and hydrogen is rapidly cooled from the melt via 
first-principles molecular dynamics simulation. The Si--H interaction 
can be  taken  into  account conveniently  within the quantum mechanical 
framework, but problems arise from the short time scale and prohibitively 
expensive scaling of computer time with the system size. The presence of 
H atoms requires a relatively small time step during MD runs limiting total 
simulation time to few picoseconds that is inadequate to  producing 
high-quality (small defect concentration), strain-free networks. Large scale realistic modeling of a-Si:H is
therefore impractical with MD alone~\cite{Robertson}.

Reverse Monte Carlo is a popular method used to efficiently generate 
structural models in agreement with experiments, as discussed 
elsewhere\cite{Biswas,gere,McG}.  RMC readily generates an ensemble of 
structures but only a subset, possibly a miniscule subset are physically 
meaningful. In RMC,  one usually remedies this problem by adding 
topological and or chemical constraints \cite{Biswas} but the constrained 
minimization is difficult within a simple Monte Carlo scheme. The way out 
of this dilemma is to identify constraints that are hierarchically most 
important along with experimental information and to merge this with 
either first-principles or a suitable classical force field. In a recent 
paper we have demonstrated that a judicious combination of these two can 
constrain the structure to evolve on the multi-dimensional energy surface 
correctly, which also is consistent with experimental data space\cite{ecmr1, ecmr2}. 

In this paper, we extend the idea of experimentally constrained molecular relaxation (ECMR) 
to model a-Si:H.  The approach is to build models by incorporating experimental information 
explicitly in the beginning, and to then relax such a starting structure to an energy minimum.  
In our first implementation of the method\cite{ecmr1} we self-consistently iterated the RMC 
and relaxation steps until convergence was obtained. In the present work, we implement this 
as a two-step process employing reverse Monte Carlo (RMC) followed by first principles molecular 
dynamics relaxation. Only one iteration of the ECMR loop is required as we discuss below. By 
construction, the models we obtain satisfy the common sense criterion that a ``good" model must 
jointly match experiments and be near to a minimum of a trusted energy functional. Their 
subsequent value (and validation) depends upon their predictive power for 
observables not employed in the construction of the models.
The RMC scheme implemented in this work can be found in Ref.~\onlinecite{Biswas}. Atoms are 
randomly packed into a  cubic supercell using periodic boundary conditions, subject to the 
constraint that no Si pair can be closer than 2 {\AA}. A quadratic cost or penalty function 
$\Pi$ is introduced in such a way that if minimized to zero, the experimental radial distribution 
function of amorphous silicon is exactly reproduced and a set of suitable constraints that 
describes the chemical and geometric properties (experimental information on the mean and 
RMS variation of bond angles, four coordination) of the system is also enforced~\cite{note2}. 
Naturally, it is impossible to find a set of coordinates that make $\Pi=0$, but it is possible 
to obtain models consistent within noise and systematic limitations (like finite-size effects) 
with the experimental information and constraints. This implies a small but finite $\Pi$.

The resulting configuration obtained from RMC relaxation 
is a 500-atom CRN having a non-uniform distribution of coordination defects, with 
dangling bonds being the dominant defect. The average bond angle of 
the CRN is found to be 109.3{\degree} with a root-mean-square (RMS) deviation of 
12.6{\degree}. 
The number of four-fold coordinated atoms is found to be 88\% 
while the remaining 12\% consists of three- and five-fold 
coordinated atoms.

The next step toward building a configuration of a-Si:H is to 
hydrogenate the RMC-generated CRN. The hydrogenation 
proceeds  following a scheme which is similar but {\em not}
identical to that proposed by Holender and Morgan~\cite{Holender}. We 
generate two configurations of a-Si:H starting from the 
500-atom CRN following two different hydrogenation schemes. 

The first scheme involves passivating only the dangling bonds while 
in the second both the dangling and floating bonds (fivefold sites) are removed. 
For the 540-atom model, the dangling bonds are identified and 
passivated by placing a H atom at a distance of 1.45 to 1.65 {\AA},  along the 
direction vector opposite to the sum 
of the three vectors connecting the central atom and its three neighbors. 
The floating bonds are kept undisturbed to minimize 
hydrogen content. The density of the a-Si:H supercell is adjusted 
to experimental value once all the dangling bonds are passivated. 
The configuration obtained after hydrogenation has defect concentration 
of approximately 2\% which is due to floating 
bonds left undisturbed during hydrogenation. 

For the 611-atom model, 
we remove all the three- and five-fold coordinated atoms. 
To remove a floating bond, we look for the nearest neighbor of 
the central atom that has a lowest coordination number. 
If a floating bond has neighbors which are all four-fold, we choose the neighbor 
which is farthest from the central atom.  We also check for pairs 
of atoms which are both five-fold coordinated and nearest neighbor 
of each other. In this case, we break the bond between two 
five-fold atoms by increasing the distance beyond the cutoff 
2.8 {\AA}. The second scheme evidently adds more hydrogen 
to the network but produces 
completely defect-free configurations.  

Once the hydrogenation is 
completed, we relax the configuration until the largest atomic 
force acting on atoms is less or equal to 0.007 eV/\AA. The relaxation 
is performed using {\sc Siesta}~\cite{Siesta1} which is a local basis, first principles 
density functional code. Because of large system size, we employ 
the Harris functional approach~\cite{Harris}, which employs a  linearized form of 
the Kohn-Sham equations. For silicon atoms we use minimal single-$\zeta$ (SZ) 
basis orbitals while for hydrogen atoms, a double-$\zeta$ basis with 
polarization orbitals (DZP) is needed to properly represent the 
H interactions\cite{note1}. The effect of basis dependence on electronic 
structure of amoprhous silicon has been discussed at length in Ref.~\onlinecite{ray-sys}.

The process of ``relax and clean" goes on until the defect concentration 
is found to be reasonably low ($ \le $ 3\%). 
For the 611-atom, the 
iteration converges after a few steps while for the 540-atom model we 
keep going until we obtain a configuration which is not only relaxed 
but also has a small defect concentration. The procedure finally yields 
a 540-atom model with 2.9\% defect concentration -- 1.9\% five-fold, 
0.3\% three-fold, and 0.7\% two-fold.  The average coordination number of 
the network is found to be 3.98. To the best of our knowledge, this 
540-atom model is possibly the largest device quality model (obtained via 
first-principles) of a-Si:H (7.4\% hydrogen) with defect concentration 
as low as 2.9\%.  The 611-atom model is found to be free from any defect 
with the exception of a single two-fold Si atom, and have 22\% of hydrogen.

\begin{figure}[htpb]
\includegraphics[width=3.24in, height=4.9in]{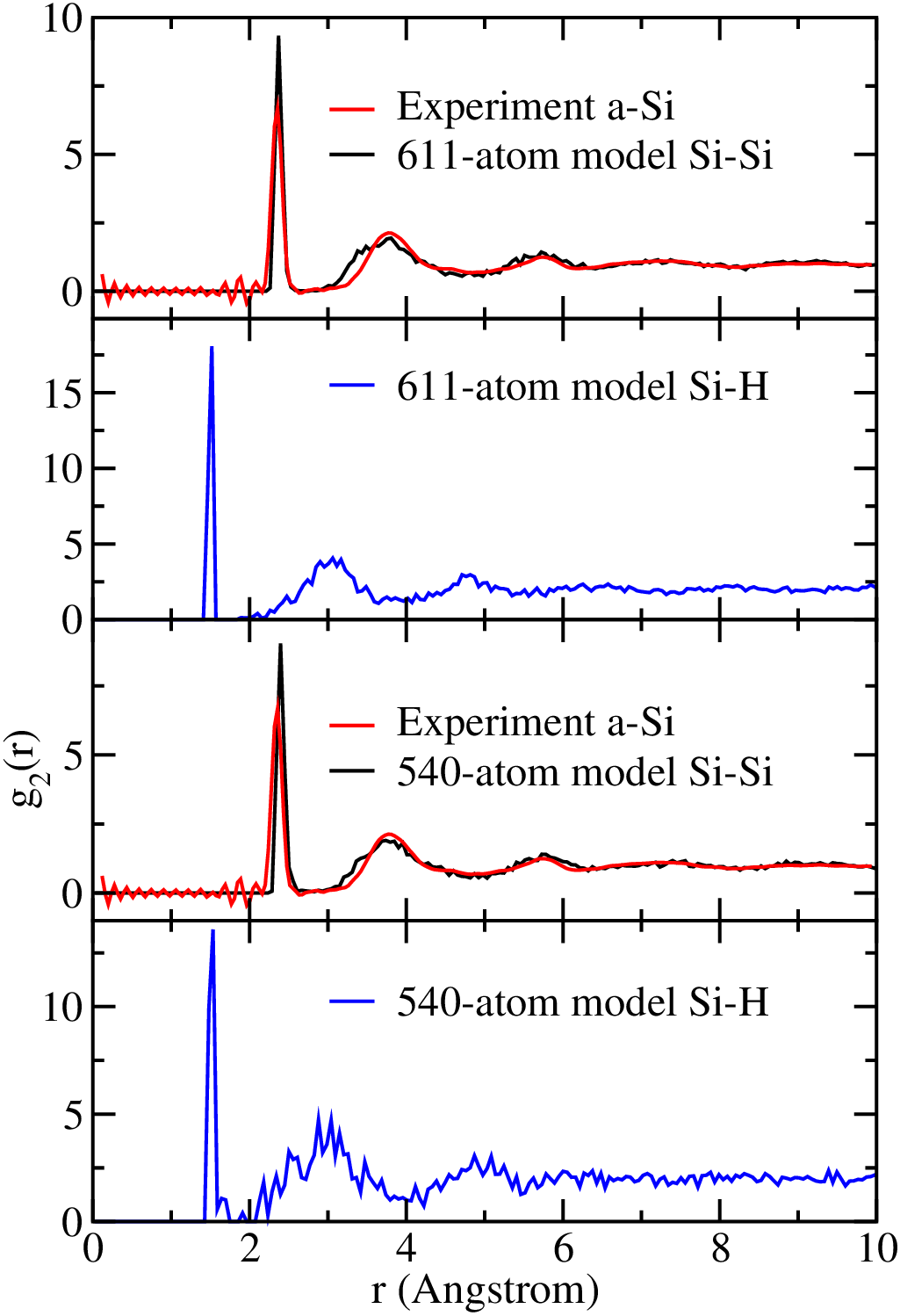}
\caption{
\label{fig1}
(Color online) The partial pair correlation functions between Si--Si and Si--H for 
the two models of a-Si:H simulated in this work. The experimental 
data for Si--Si partial are also shown in figure (red/gray) for comparison 
with our results. 
}
\end{figure}

In Fig.~\ref{fig1} we present the partial pair correlation functions.  
along with experimental data reported by Laaziri et al~\cite{laaziri}.  Although 
the information on Si--Si structure factor has been ``built into" our 
starting RMC-generated CRN, it is important to ensure that the hydrogenation 
and relaxation does not introduce any changes in Si-Si pair correlation. 
It is clear from the Fig.~\ref{fig1} that the Si--Si pair correlation from 
our model agrees closely with that obtained from experiment~\cite{laaziri}. 
It is to be noted that the experimental data are for pure amorphous silicon, 
whereas Si-Si partial pair correlation function in Fig.~\ref{fig1} is computed 
in hydrogenated environment. This possibly explains the difference in peak height
of our result with the experimental data and the results obtained in Ref.~\onlinecite{Bernstein}. 
Apart from this slight deviation, our result matches with the experimental data 
reasonably good.  This indicates that as the system evolves via ECMR, 
it moves progressively toward a more favorable configuration consistent with 
both total energy and experimental data. 

\begin{figure}[htpb]
\includegraphics[width=3.2in, height=4.2in]{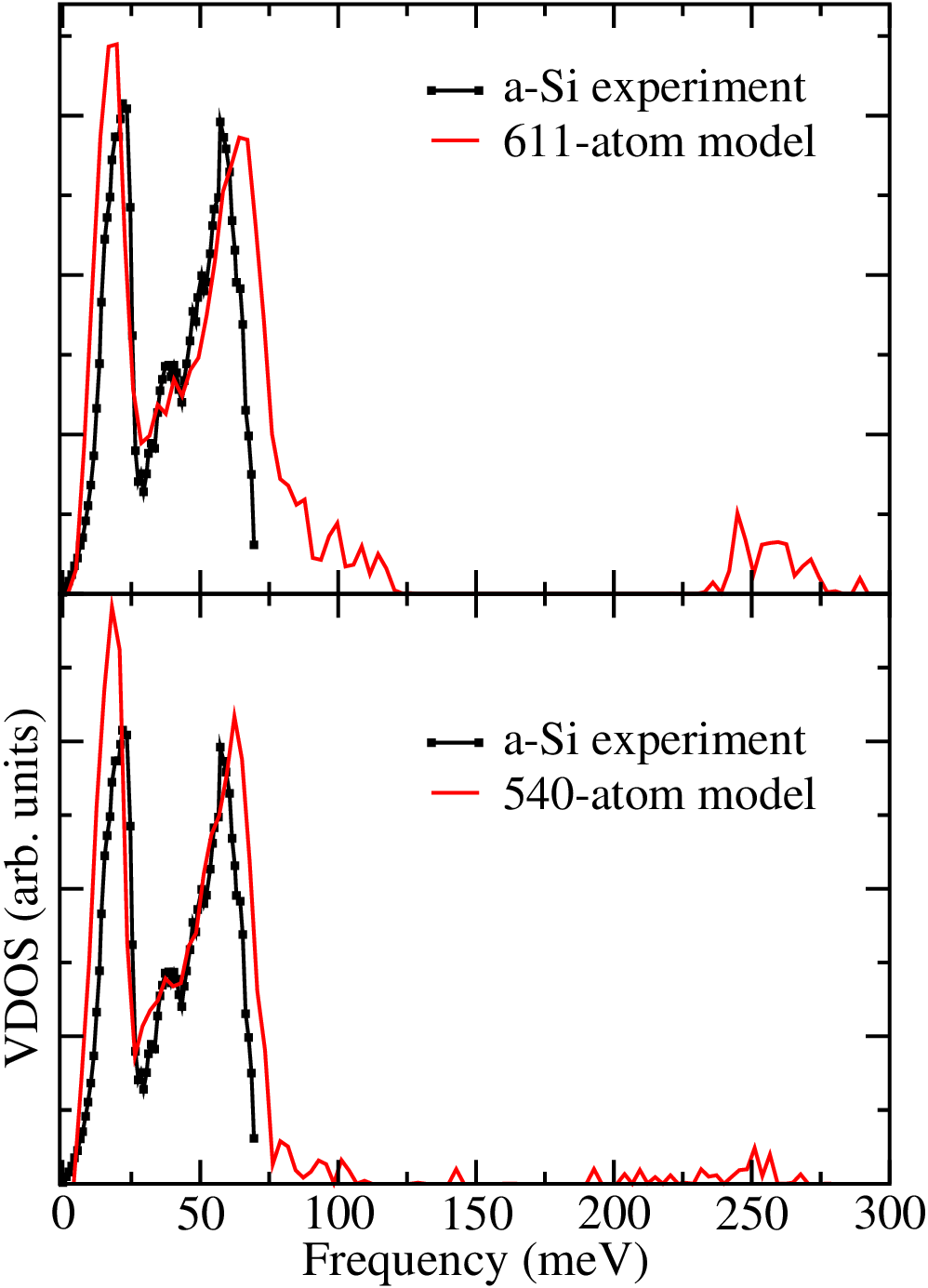}
\caption{\label{fig2}
(Color online)
Vibrational density of states for the two models of a-Si:H.
The upper and the lower panel stands for 611- and 540-atom (red/gray)
model respectively. The experimental data for (unhydrogenated) amorphous Si from Neutron 
diffraction are also plotted for comparison (black). 
}
\end{figure}

In Fig.~\ref{fig2}, the vibrational density of states (VDOS) is plotted for 
both models. The VDOS is computed by diagonalizing the dynamical matrix, which
is obtained from the force constant matrix. The force constant matrix is approximated with
finite differences obtained from $6N$ calculations of forces on all atoms for suitably chosen small 
displacements from the equilibrium conformation.

\begin{figure}[htpb]
\includegraphics[width=3.25in, height=2.8in]{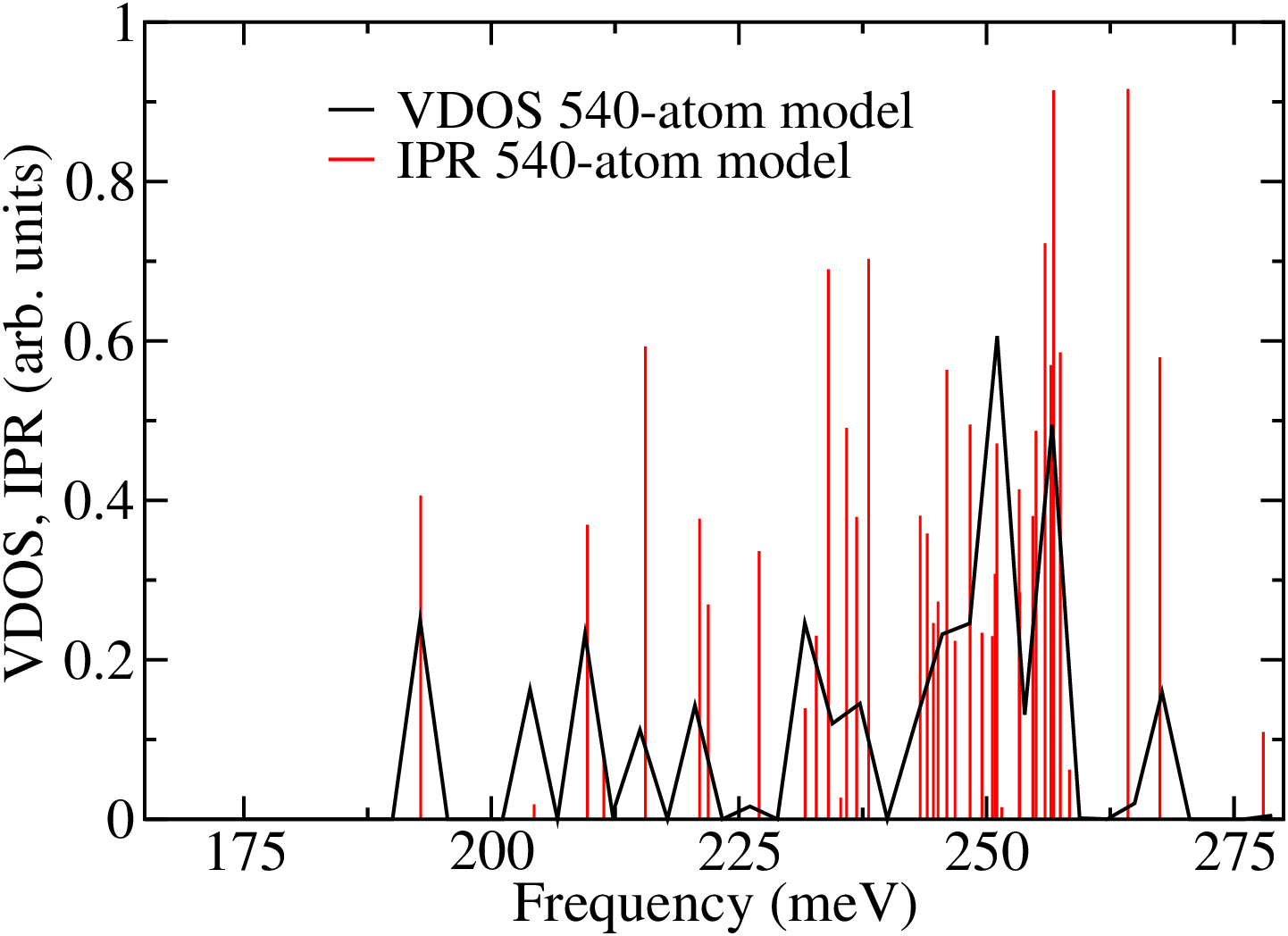}
\caption{\label{fig3}
(Color online) The inverse participation ratio (red/gray) of vibrational modes along 
with their density of states (black) for 540-atom model. The localized nature 
of the high frequency band is clear from the large inverse participation ratio. The VDOS 
is shown as a guide to the eyes and is scaled for plotting. 
}
\end{figure}

Since the eigenvalues of the dynamical matrix are given by the square of the angular frequency; 
one can construct the density distribution of the frequencies from a knowledge of the eigenvalues. 
The acoustic and the optical peaks appear at correct position (22.0 and 61.0 meV 
respectively) which is in good agreement with the experiment data reported by 
Kamitakahara et al.~\cite{kam} and is also shown in the figure. Both the models 
show an excess of high frequency modes which are associated with hydrogen atoms. 
Such high frequency modes have been observed in the experimental vibrational spectrum 
of hydrogenated a-Si~\cite{Voort}.  Of particular importance is the vibrational modes around 
250 meV ($\approx$ 2000 cm$^{-1}$).  Rella et al. have studied the localization 
of Si-H stretch vibration in amorphous silicon using vibrational transient grating 
measurement, and observed highly localized modes around 2000 
cm$^{-1}$~\cite{Rella, Voort}. The presence of this high frequency vibrational 
band adds further credibility to our model. 
To characterize the spatial extent of the vibrational states we have calculated 
the inverse participation ratio (IPR) of the states in this band obtained from the 
dynamical matrix. The IPR for a normalized state j is defined as: 

$$
I = \sum_{i=1}^N ({\Phi^j_i}\, . \, {\Phi^j_i})^2 
$$

and provides a measure of the inverse of the number of sites associated with a 
state. Here $\Phi^j$ is a normal mode eigenvector and N is the number of atoms (molecules 
in a non-monoatomic system) in the 
system. For an ideally localized state, only one atom (molecule) contributes to the vibration giving 
$I = 1$ whereas for a uniformly extended state we have $I = 1/N$. 
In Fig.~\ref{fig3} we have plotted the inverse participation ratio along with the 
vibrational density of states for the 540-atom model. The localized nature of the 
high frequency vibrational band is clearly visible from the plot.  Our preliminary study 
indicates that this vibrational band is due to isolated localized states of Si-H 
vibrations, and the vibrational modes correspond to frequency around 2000 cm$^{-1}$ 
(250 meV in the Fig.~\ref{fig3}) show very much like the character of a Si-H stretching 
mode. However, a complete analysis of the nature of the vibration requires a 
self-consistent relaxation of the structure 
and an accurate determination of vibrational frequencies and eigenvectors. 
Investigations are presently being carried out to study the nature of vibration 
modes in this band. 

\begin{figure}[htpb]
\includegraphics[width=3.25in, height=3.8in]{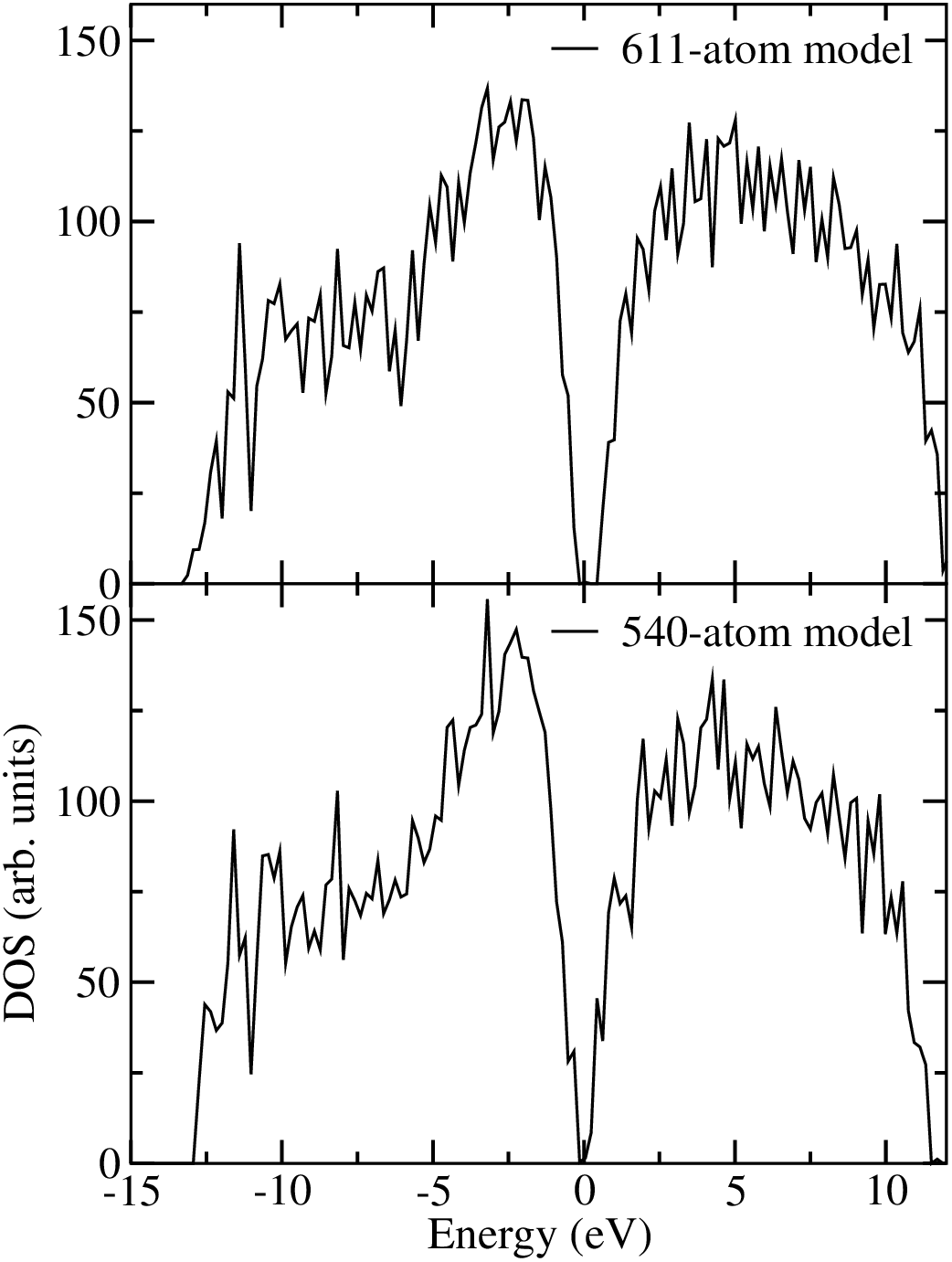}
\caption{
\label{fig4}
The electronic density of states for the two models using Harris functional and local 
density approximation (LDA). The Hamiltonians for the models are 
constructed using single-$\zeta$ and double-$\zeta$-polarized basis for Si and H 
atoms respectively.
}
\end{figure}

Finally,  in Fig.~\ref{fig4} we have plotted the electronic density of states 
of the models obtained from {\sc Siesta} using the local density approximation. 
Both the models show the presence of a clean gap in the spectrum. This electronic 
gap is very sensitive to the presence of defect states and other imperfections 
present in the model. For the 540-atom model, the electronic gap is slightly 
narrower than its 611-atom counterpart which is quite expected because of 2.9\% 
concentration of defect compared to a single defect in the 611-atom model. There are no 
localized mid-gap states and a few band-tail states in the 540-atom model. These tail 
states are generally considered to be associated with the presence of disorder in 
bond length and bond angle distribution. Recent work by Bernstein et al. have 
indicated that the width of the bond length distribution can affect the density 
of states in near the band edge region~\cite{Bernstein}. 
We believe that further tuning of electronic density of states is possible via prolonged 
relaxation using our experimental constrained molecular relaxation 
(ECMR) method~\cite{ecmr1} which we have recently developed and applied 
to glassy GeSe$_2$.

In conclusion, we apply ECMR to hydrogenated amorphous silicon. The novelty of this 
approach is its simultaneous use of both experimental information and the information
implicit in an accurate energy functional. We apply this method two produce two 
large models of a-Si:H. The structural, vibrational, 
and electronic properties of the models agree very well with experiments. We show that the 
method is fast and is capable of generating device quality a-Si:H models with low defect 
concentration. A comparison with existing models (cf.~Table I) obtained by other 
researchers clearly reveals the fact that the method benefits from both static and 
dynamic approaches -- it can handle a large number of atoms (a characteristic feature of 
static approach) and at the same time producing highly relaxed configuration with average 
force on atoms as small as few meV/{\AA}.  A remarkable feature of the models is that they produce 
a high frequency localized vibrational band around 250 meV which has been observed 
experimentally. Analysis of eigenvectors corresponding to this and other nearby
modes suggests the models are localized. The models generated in this work 
can be used as a starting point for an in-depth study of properties of H in amorphous 
silicon using a full self-consistent calculations.  The approach is very general and has the 
potential to impact other complex materials.  

\begin{center}
\begin{table}[htbp]
\caption{\label{tab2} Summary of existing bulk models ($ > $ 100 atoms) of a-Si:H }
\begin{ruledtabular}
\begin{tabular}{lllcc}
Authors (year) & Model size & \% of H & Method \\
\hline
ML (1991)\cite{Normand} & 270 & 16.0 & Static\\
HMJ (1993)\cite{Holender} & 2025  & 23.0 & Static+dynamic\\
TA\cite{Tuttle} (1996) & 242  & 11.0 & dynamic \\
KUF\cite{Klein} (1999) & 128  & 6.25 & dynamic  \\
Present work (2007) & 540 (611) & 7.4 (22)  & ECMR \\
\end{tabular}
\end{ruledtabular}
\end{table}
\end{center}

DAD thanks the NSF for support under grants DMR 0605890 and 0600073, and the ARO under 
MURI W911-NF-06-2-0026. PB acknowledges the support of the University of Southern 
Mississippi under Grant No.~DE00945. We thank one of the referees for pointing references 
\onlinecite{Bernstein}, \onlinecite{Voort} and \onlinecite{Rella}.

\newpage 

\newpage 

\newpage 

\newpage 


\begin{thebibliography}{99}
\bibitem{note2}
The zero-point motion of the nuclei is not taken into account in our approach. At room 
temperature, the effect on Si atoms can be neglected other than an additional broadening 
of the first peak of the radial distribution function. The effect on H atoms may be non-trivial, 
which can be addressed via path integral Quantum Monte Carlo at high temperature 
or via harmonic approximation at low temperature as discussed in Refs.~\onlinecite{herreo} 
and ~\onlinecite{feldman} respectively.
\bibitem{note1}
Because of the heavy computational nature of the problem, we have not studied the role of 
full self-consistency and the basis dependence of Si-H bonding in our present work.  Such 
issues might be important in understanding the electronic structure and dynamics of H atoms
and will be addressed in a separate communication. For a discussion on a-Si, see 
Ref.~\onlinecite{ray-sys}.
\end{thebibliography}
\end{document}